\documentclass[prd,twocolumn,showpacs,preprintnumbers,amsmath,amssymb,superscriptaddress,nofootinbib]{revtex4-1}

\usepackage[T1]{fontenc}
\usepackage{ae,aecompl}

\usepackage{graphicx}	
\usepackage{amsmath}	
\usepackage{amssymb}	

\usepackage[colorlinks=true,
  linkcolor=black,
  anchorcolor=blue,
  citecolor=blue,
  filecolor=blue,
  menucolor=blue,
  urlcolor=blue]{hyperref} 

\usepackage{aas_macros}
\usepackage{amsfonts}
\usepackage{booktabs}
\usepackage{bbm}
\usepackage{graphicx}
\usepackage{multirow}
\usepackage{subcaption}
\usepackage[separate-uncertainty=true]{siunitx}

\renewcommand{\citep}{\cite}
\renewcommand{\citet}{\cite}

\newcommand{\chisq}{\ensuremath{\chi^2}}
\newcommand{\dlum}{\ensuremath{d_\text{L}}}

\newcommand{\dang}{\ensuremath{d_\text{A}}}

\newcommand{\hMpc}[1]{\ensuremath{#1\,h^{-1}\mathrm{Mpc}}}
\newcommand{\lcdm}{\texorpdfstring{$\Lambda$}{L}CDM}

\newcommand{\Ok}{\ensuremath{\mathcal{O}_k}}
\newcommand{\Om}{\ensuremath{Om}}
\newcommand{\Omm}{\ensuremath{\Omega_\text{m}}}
\newcommand{\Omb}{\ensuremath{\Omega_\text{b}}}
\newcommand{\Omde}{\ensuremath{\Omega_\text{de}}}

\newcommand{\Omk}{\ensuremath{\Omega_k}}
\newcommand{\Omr}{\ensuremath{\Omega_\text{r}}}
\newcommand{\one}{\ensuremath{\mathbbm{1}}}
\newcommand{\Hord}{\ensuremath{H_0\rd}}
\newcommand{\vect}[1]{\ensuremath{\boldsymbol{#1}}}
\newcommand{\tens}[1]{\ensuremath{\mathbf{#1}}}
\newcommand{\rd}{\ensuremath{r_\text{d}}} 
\newcommand{\diff} {\ensuremath{\mathrm{d}}} 
 
\newcommand{\deriv}[2]{\ensuremath{\frac{\diff {#1}}{\diff {#2}}}}

\newcommand{\mean}[1]{\ensuremath{\left\langle #1 \right\rangle}}

\newcommand{\DD}{\ensuremath{\mathcal{D}}}

\newcommand{\chsq}{\ensuremath{\chi^2}}
\newcommand{\fseight}{\ensuremath {f\sigma_8}}

\begin{document}
\label{firstpage}

\title{Falsifying \lcdm: Model-independent tests of the
concordance model with eBOSS DR14Q and Pantheon}

\author{Arman Shafieloo}
\email{shafieloo@kasi.re.kr}
\affiliation{Korea Astronomy and Space Science Institute, Yuseong-gu, Daedeok-daero 776, Daejeon 34055, Korea}
\affiliation{University of  Science and Technology,  Yuseong-gu 217 Gajeong-ro, Daejeon 34113, Korea}

\author{Benjamin L'Huillier}
\email{benjamin@kasi.re.kr}
\affiliation{Korea Astronomy and Space Science Institute, Yuseong-gu, Daedeok-daero 776, Daejeon 34055, Korea}

\author{Alexei A.~Starobinsky}
\email{alstar@landau.ac.ru}
\affiliation{L. D. Landau Institute for Theoretical Physics RAS, Moscow 119334, Russia}
\affiliation{National Research  University Higher  School of  Economics, Moscow 101000, Russia }



\begin{abstract}
We combine model-independent reconstructions of the expansion history
from the latest Pantheon supernovae distance modulus compilation and
measurements from  baryon acoustic oscillation to  test some important
aspects of the  concordance model of cosmology namely  the FLRW metric
and flatness of spatial curvature.
We  then  use the  reconstructed  expansion  histories to  fit  growth
measurement   from  redshift-space   distortion   and  obtain   
constraints on $(\Omm,\gamma,\sigma_8)$ in a model independent manner.
Our results show consistency with  a spatially flat FLRW Universe with
general  relativity  to  govern  the  perturbation  in  the  structure
formation and the cosmological constant as dark energy.
However,  we  can also  see  some  hints  of tension  among  different
observations within  the context of  the concordance model  related to
high  redshift observations  ($z>1$) of  the expansion  history.  This
supports  earlier   findings  of   \citet[][]{2014ApJ...793L..40S, 2017NatAs...1..627Z}  and  highlights the  importance  of
precise measurement  of expansion history  and growth of  structure at
high redshifts.
\end{abstract}


\maketitle



\section{Introduction}

The  concordance model  of cosmology  is based  on Einstein's  general
theory of relativity (GR), which  enabled us to build a theory of the
Universe that is testable and can be falsified.
The concordance flat $\Lambda$CDM model, which  is based on GR and the
assumptions of isotropy and homogeneity of the Universe, has been very
successful at explaining various astronomical observations  from a very early epoch (at least, from the Big-Bang nucleosynthesis time).
This predictive model explains the  dynamics of the Universe with only
6 free parameters. 
$\Omega_\text{b}$  and $\Omega_\text{dm}$  (baryonic  and dark  matter
densities) are the matter parameters.
Assuming a flat universe  and cosmological constant being responsible  for late time
acceleration      of     the      Universe,     we      can     derive
$\Omega_{\Lambda}=1-(\Omega_\text{b}+\Omega_\text{dm})$. 
$\tau$  representing  the  epoch   of  reionization,  $H_0$  the  Hubble
parameter, $n_s$  the spectral  index of  the primordial  spectrum and
$A_s$ the overall amplitude of the primordial spectrum are the other 4
parameters of this model.
Out of  these parameters, the  first four  dictate the dynamic  of the
Universe and  the other  two represent  the initial  condition through
the primordial fluctuations given by
$P_R(k)=A_s\left(\tfrac{k}{k_*}\right)^{n_s-1}$,  where  $k_*$ is  the
pivot point.  
Having  the form  of  the primordial  fluctuations  and the  expansion
history of the  Universe one can determine the growth  of structure for
this  model on  linear  scales following  the linearised  perturbation
equation and also  run $N$-body simulations to study  the small scales
and non-linear regime.
Despite the simplicity of the model, most astronomical observations are
in great agreement with the concordance model and so far there has not
been any strong observational evidence against it \citep[e.g.,][]{2016A&A...594A..13P, 2017MNRAS.470.2617A, 2018ApJ...859..101S}.  
In this paper we test some  important aspects of the concordance model
of cosmology in light of the most recent cosmological observations in a
model-independent manner.
At the background level, we derive the $H_0 r_d$
parameter, test dark  energy as the cosmological constant  $\Lambda$, the FLRW
metric and the  flatness of the Universe. 
At the perturbation level, we then use model independent  reconstruction of the expansion history
from supernovae  data to fit  growth of  structure data and  put model
independent   constraints    on  some key   cosmological    parameters, namely 
$\Omm,\gamma$, and $\sigma_8$.  
In \S~\ref{sec:background}  we describe  the background  expansion and
our tests  on $\Lambda$ dark energy,  FLRW metric and flatness  of the
spatial curvature.  Analysis on  the growth  of structure  and testing
general theory of relativity are presented in \S~\ref{sec:growth}, and
our conclusions are drawn in \S~\ref{sec:ccl}.

\section{Background Expansion: Testing \texorpdfstring{$\Lambda$}{Lambda}, the FLRW Metric, and the Curvature}

\label{sec:background}

At the background level, it is possible to test dark energy as $\Lambda$,
the FLRW metric, and the curvature of the Universe.  
In a FLRW  universe with a dark energy component  of equation of state
$w(z)$, the luminosity distance can be written for any curvature \Omk\ 
\begin{align}
  \dlum(z) & = \frac c {H_0} (1+z) \DD(z), \intertext{where}
  \DD(z)  &= \frac  1  {\sqrt{-\Omk}} \sin\left(\sqrt{-\Omk}  \int_0^z
  \frac{\diff x}{h(x)}\right) 
  \intertext{is the dimensionless comoving distance, and}
  h^2(z) & = \left(\frac{H(z)}{H_0}\right)^2 = \Omm (1+z)^3 + \Omk(1+z)^2 \nonumber\\
  & + (1-\Omm-\Omk) \exp{\left(3\int_0^z \frac{1+w(x)}{1+x} 
  \diff   x\right)}
\end{align}
is the expansion  history. Having different observables  of the cosmic
distances  and  expansion history  one  can  then  introduce  novel
approaches to  examine the FLRW  metric, flatness of the  Universe and
$\Lambda$    dark    energy    in     a    model-dependent \citep[e.g.,][]{2013PhLB..723....1F} 
or    independent    way
\citep{2008PhRvL.101a1301C, 2008PhRvD..78j3502S, 2014PhRvD..90b3012S, 2017JCAP...01..015L, 2017NatAs...1..627Z, 2018PhRvD..97h3510M,2017JCAP...11..032M}.  
Note that one can also test the metric and the curvature using gravitational lensing 
\citep[e.g.][]{2015PhRvL.115j1301R, 2018JCAP...03..041D} or cosmic parallaxes \citep{2014JCAP...03..035R}.

\subsection{Model-independent   reconstruction  of   the  expansion
  history from the Pantheon compilation}

In order to reconstruct the $D(z)$, $D'(z)$ and $h(z)$ at any given
redshift, we apply the iterative smoothing method
\citep{2006MNRAS.366.1081S, 2007MNRAS.380.1573S, 2010PhRvD..81h3537S,
  2017JCAP...01..015L} to the the latest compilation of supernovae 
distance modulus \citep[Pantheon,][]{2018ApJ...859..101S}. Pantheon is
the latest compilation of 1048 SNIa, extending previous compilations
with confirmed SNIa from the Pan-STARRS1 survey.

The method of smoothing is a fully model independent approach to reconstruct the $D(z)$ relation directly from the supernova data, without assuming any particular model or a parametric form. The only parameter used in the smoothing method is the smoothing width $\Delta$, which is constrained only by the quality and quantity of the data. The smoothing method is an iterative procedure with each iteration providing a better fit to the data. It has been discussed and shown that the final reconstructed results are independent of the assumed initial guess~\citep{2006MNRAS.366.1081S, 2007MNRAS.380.1573S, 2010PhRvD..81h3537S}. In our analysis we start the smoothing procedure from various arbitrary choices of the initial guess models and while their final results converge to the same reconstruction, we select within the process, a non-exhaustive samples of the reconstructions that have a $\chi^2$ better than the best fit $\Lambda$CDM model. In ~\citep{2010PhRvD..81h3537S} the method of smoothing was modified to incorporate the data uncertainties and hence making the approach error-sensitive. However, the formalism in ~\citep{2010PhRvD..81h3537S} could take in to account only the diagonal terms of the error matrix. While the quality of the data is improving continuously and non-diagonal terms of the covariance matrices can play an important role in likelihood estimations, in this work we modify the smoothing method further by incorporating the whole covariance matrix of the data in to the smoothing procedure. While this improvement might look like a minor modification, it is in fact a very important step to make this model independent reconstruction approach complete and comprehensive to deal with highly correlated data.

In order to take into account the non-diagonal terms of the covariance
matrix, we modified the method in the following way. 
Starting with some initial guess $\hat\mu_0$, we iteratively calculate
the reconstructed $\hat\mu_{n+1}$ at iteration $n+1$:
%
\begin{align}
  \hat\mu_{n+1}(z) & = \hat\mu_n(z) + \frac{\vect{\delta\mu}_n^\mathrm{T} \cdot
  \tens{C}_\text{SN}^{-1} \cdot \vect{W}(z)}  
  {\one^T \cdot \tens{C}_\text{SN}^{-1} \cdot \vect{W}(z)},\label{eq:smoothing1}
  \intertext{where the weight \vect W and residual \vect{\delta \mu_\mathrm{n}} are defined as}
  \vect{W}_i(z) & =  {\exp{\left(- \frac{\ln^2\left(\frac
      {1+z}{1+z_i}\right)}{2\Delta^2}\right)}}\label{eq:smoothing2}\\
  \vect{\delta\mu}_n|_i & = \mu_i-\hat\mu_n(z_i),\label{eq:smoothing3}\\
\one^T   & = (1,\dots,1),
\end{align}
 and $\tens{C}_\text{SN}$ is the covariance
matrix of the data (in our case, Pantheon data). 
In case of uncorrelated data ($C_{ij} = \delta_{ij} \sigma_i^2)$), we 
recover   the   formula introduced in ~\citep{2010PhRvD..81h3537S}  used recently in  \citet{2017JCAP...01..015L}.

The $\chi^2$ of the reconstruction $\hat\mu_n(z)$ is then defined as
\begin{align}
  \chi^2_n  &  =  \vect{\delta\mu}_n^\text{T}  \cdot  \tens{C}_\text{SN}^{-1}
  \cdot \vect{\delta\mu}_n, \label{eq:smoothing4}
\end{align}
and in this work we only consider reconstructions with
$\chi^2<\chi^2_\text{\lcdm\ best-fit}$. 

The result  of the smoothing  procedure is thus $H_0  \hat\dlum_n(z) =
10^{(\hat\mu_n-5)/5}$. 
Under  the assumption  of a  flat Universe,  we can  obtain $h_n(z)  =
1/(\diff \DD_n(z)/\diff z)$. \\

We should clarify here that our selected reconstructions of the expansion history from the iterative smoothing method are not posterior samples within a Bayesian framework. We in fact obtain a  non-exhaustive sample of  plausible expansion histories, directly reconstructed by supernova data and with no model assumption, which all give a better \chsq\ to the  Pantheon data than the best-fit \lcdm\ model. This  enables us  to  explore regions  of the  physical  space of  the expansion history beyond the flexibility of the concordance model (or other parametric functional forms) that can fit the data reasonably well.  
Note that the formalism given in this paper for the method of smoothing is self-contained and has all needed information. Equations \eqref{eq:smoothing1} to \eqref{eq:smoothing4} contain the full formalism of the iterative smoothing method including the full covariance matrix of the data, which is now simply written in a matricial way (which is more compact). 
However, for more details and better understanding of the method one can follow the given references.

\subsection{BAO measurements of cosmic distances and expansion history}

The radial  mode of the  BAO measures $H(z)\rd$, while  the transverse
modes provide $\dang(z)/\rd$, where 
\begin{align}
\rd &  = \frac  c {\sqrt  3} \int_0^{1/(1+z_\text{drag})}  
\frac {\diff  a} {a^2 H(a) \sqrt{1+\frac{3\Omb}{4\Omr} a}} 
\end{align}
is the sound horizon at the drag epoch $z_\text{drag}$.
We combined  the Baryon  Oscillation Spectroscopic Survey  (BOSS) DR12
consensus  values  \citep{2017MNRAS.470.2617A} and  the  extended-BOSS
(eBOSS) DR14Q measurements \citep{2018arXiv180103043Z}. 
We note that both BOSS    DR12    and    eBOSS     DR14Q    provide
$H(z)\rd/r_\text{d,fid}$    and   $\dang(z)r_\text{d,fid}/\rd$    with
$r_\text{d,fid}  = \SI{147.78}{Mpc}$.
We also  include the Dark Energy  Survey DR1 (DES DR1)  measurement of
$\dang/\rd$ at $z=0.81$ \citep{2017arXiv171206209T}.
We use these BAO data along  with our reconstructions of the expansion
history from supernova data as two independent sets of observations to
test some key aspects of the concordance model.

\begin{figure}
  \centering
  \includegraphics[width=\columnwidth]{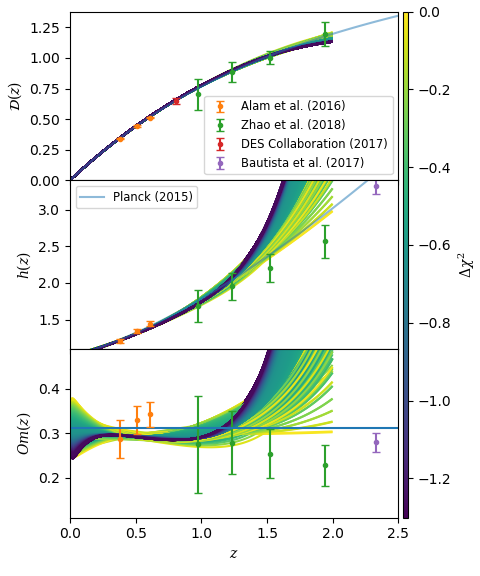}
  \caption{\label{fig:bao}
    BAO      data     points      normalized     by      \Hord\     from
    \citep{2016A&A...594A..13P} best fit $\Lambda$CDM model.  
    The solid lines are the reconstructed expansion histories
    from the  Pantheon data which  are fully model independent,  and the
    purple line is the prediction from 
    \citep{2016A&A...594A..13P}    for    the    best-fit    concordance
    $\Lambda$CDM model.  
    They are color-coded by their $\Delta\chi^2$ with respect to the
    best-fit \lcdm\ model, with earlier iterations having less
    negative $\delta\chi^2$ (yellow), and later iterations more
    negative $\Delta\chi^2$ (dark blue). 
  }
\end{figure}

\subsection{Testing \texorpdfstring{$\Lambda$}{Lambda} Dark Energy}

The  solid  black  lines  in  Fig.~\ref{fig:bao}  show  the  different
reconstructed $\DD(z)$,  $h(z) = 1/\DD'(z)$ and  $Om(z)$ from Pantheon
supernovae    compilation     where    $Om(z)$    is     defined    as
\citep{2008PhRvD..78j3502S}: 
\begin{align}
Om(z) = \frac{h^2-1}{(1+z)^3-1}
\end{align}

We  also show  in Fig.~\ref{fig:bao}  the  BAO data  points for  these
quantities. 
Since the  BAO measure  $H(z)\rd$ and $\dang(z)/\rd$,  to have  a good
sense of  comparison within the  context of the concordance  model, we
normalize them  by \Hord\ from Planck  2015 (TTTEEE+LowP+Lensing) best
fit $\Lambda$CDM  model, and  show on  the top  panel $\DD(z)  = (1+z)
\Hord \dang(z)/(c \rd)$, in the middle panel $h(z) = H(z)\rd / H_0\rd$
and the corresponding $\Om(z)$ on the bottom panel. 
The magenta  solid line shows  the corresponding $\DD(z)$,  $h(z)$ and
$Om(z)$ for the best-fit Planck 2015 Flat-\lcdm\ model.

While the reconstructed  expansion history $h(z)$ from  SNIa are fully
consistent with  the BAO data  points at low redshifts  ($z\leq 1.2$),
some tension seems to arise at higher redshifts ($z\geq 1.5$) where the
reconstructed  expansion histories  from  the BAO  data suggest  lower
$h(z)$ with respect to the best fit $\Lambda$CDM model from Planck.
While the errorbars are still quite large, the BAO data seem to follow
the same trend  in suggesting lower values of $h(z)$  (with respect to
the best fit $\Lambda$CDM model from Planck) at high redshifts.
For illustration purpose  we also show the  measurement of $h(z=2.33)$
from the Lyman-$\alpha$ forest \citep{2017A&A...603A..12B} which seems
to  agree with  other BAO  data  points suggesting  lower $h(z)$  with
respect to  Planck best  fit $\Lambda$CDM model,  although we  did not
include this data  point in our analysis since the  supernovae data do
not reach such a high redshift.
This data point is consistent with the previous result from SDSS III \citep{2015A&A...574A..59D}.
This tension  is also  visible clearly looking  at the $\Om$  diagnostic in
bottom plot of Fig.~\ref{fig:bao}, which is also consistent with the finding of \citep{2014ApJ...793L..40S}.
If dark energy is a cosmological constant (and if the Universe is flat), the $\Om$ diagnostic should
be constant in redshift.
Therefore,  having   different  values  from   different  observations
suggests  some tension  among the  data  within the  framework of  the
concordance model.

Meanwhile, the comoving distances $\DD(z)$ from BAO and SNIa are fully
consistent together and with the best-fit Planck cosmology.
Combining  these  results  of  the comoving  distances  and  expansion
histories may  show some  inconsistency with flatness  as we  will see
later in this work.

\subsection{Estimating \texorpdfstring{$H_0\rd$}{H0rd}}

\label{sec:H0rd}

Ref.~\citet{2017JCAP...01..015L} estimated $H_0\rd$  in a model-independent
way by combining BAO measurements and reconstructions of the expansion
history from supernovae.
{\Hord\ is an important parameter combining physics of the early (sound horizon
at the drag epoch) and late Universe (expansion rate).} 
For each reconstruction $n$, we can calculate \Hord\ in two different ways
\begin{subequations}
  \begin{align}
    \label{eq:H0rd}
    H_0\rd|_{\dang,n} & = \frac c {1+z} \DD_n(z) \frac{\rd}{\dang(z)}\\
    H_0\rd|_{H,n} & = \frac{H(z)\rd}{h_n(z)},
  \end{align}
\end{subequations}
and their associated errors
\begin{subequations}
  \begin{align}
    \label{eq:sH0rddA}
    \sigma_{H_0\rd}|_{\dang,n}    &   =    \frac   c    {1+z}   \DD_n(z)
    \frac{\sigma_{\dang/\rd}(z)}{(\dang(z)/\rd)^2}\\
        \label{eq:sH0rdH}
    \sigma_{H_0\rd}|_{H,n} & = \frac{\sigma_{H\rd}(z)}{h_n(z)},
  \end{align}
\end{subequations}
where, assuming a flat-FLRW universe, $h(z) = 1/\DD'(z)$.

\begin{figure}
\centering
 \includegraphics[width=\columnwidth]{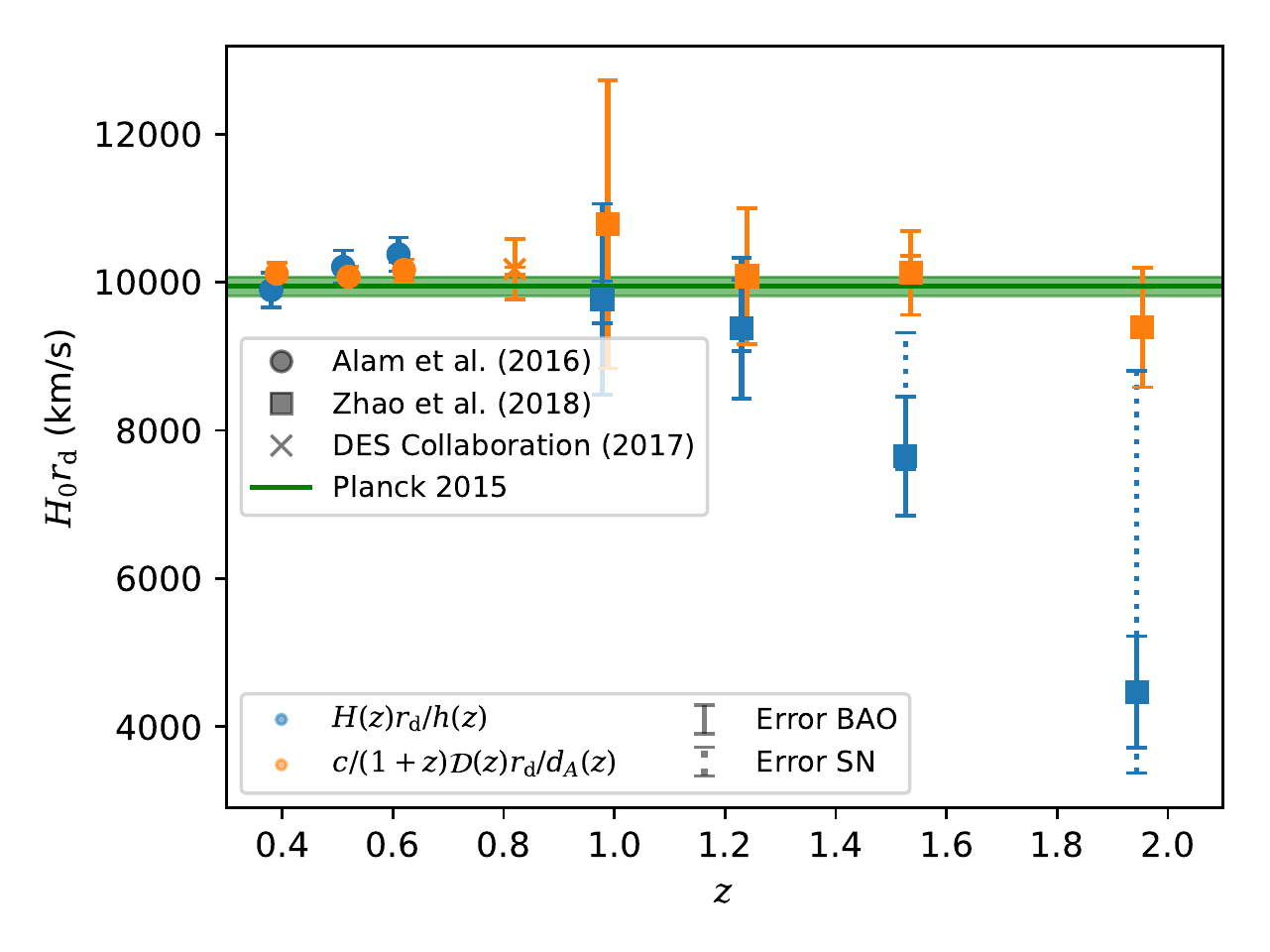}
  \caption{\label{fig:H0rd}Model-independent       measurement      of
    \Hord\ estimated at the different BAO data points.
  The dotted error-bars show the range of possible central values from
  different  reconstructions (SN  error), while  the solid  error bars
  show the uncertainty on the central value (BAO error).}
\end{figure}

Fig.~\ref{fig:H0rd} shows our estimation  of $H_0\rd$ at the different
BAO  data points  for  the  two estimations.  In  green  is shown  the
$\Lambda$CDM value from Planck 2015 \citet{2016A&A...594A..13P}. 
We can then define two error-bars. 
The first one is the error due to the supernova. 
At fixed redshift,  we define $\mean{\Hord}_X$ as the  median over all
reconstructions for method $X \in \{\dang,H\}$.  
We  can then  define the  upper  and lower  limit as  the minimal  and
maximal values of ${\Hord}|_{X,n}$.  
This error-bar is shown as a dashed line in Fig.~\ref{fig:H0rd}. 
The   second  error   is   due   to  the   uncertainty   on  the   BAO
(equations~\eqref{eq:sH0rddA} and~\eqref{eq:sH0rdH}), and is the uncertainty of the central
value for a given reconstruction $n$.  
For  each  reconstruction  $n$  and  method  $X$,  we  have  an  error
${\sigma_{\Hord}}|_{X,n}$.
They are of  the same order for each reconstruction,  so we define the
final BAO error as the maximum value over all reconstructions.  
This error-bar is shown as a solid error-bar in Fig.~\ref{fig:H0rd}.

For  the first  method (in  orange), the  measurements of  \Hord\ from
combination of supernova  and SDSS BAO data are  fully consistent with
Planck.
The  DES  data point,  also  using  the  transverse  BAO mode,  is  an
independent confirmation at intermediate redshift.  
However,  for   the  second  method,   while  at  low   redshift,  the
measurements are  consistent with  Planck, the  eBOSS data  points are
systematically lower than the Planck best-fit at $z\geq 1.2$ while the
errorbars become very large at this range.  
This can be understood by the following remarks. 

The first method  yields very consistent results thanks to  the use of
the transverse  BAO mode, which  has smaller error-bars,  coupled with
direct reconstructions $\DD(z)$ which do not use derivative.

The second  method however,  uses the line-of-sight  mode of  the BAO,
together with $h(z)$ from supernovae data which is a derivative. Since
the Pantheon data become scarce at $z\geq 1$, the estimation of $h(z)$
becomes less precise at this range having large error-bars.
Combination of these  two results to large  uncertainties for $H_0\rd$
from the second method.
On the other  hand, it can be seen from  Fig.~\ref{fig:bao} that while
$h(z)$ from  SNIa are  higher than the  best-fit Planck  \lcdm\ model,
$h(z)$ from the  BAO (scaled with best fit  Planck $\Lambda$CDM model)
are actually lower.
This  explains  the  lower  values  of  $H(z)\rd/h(z)$  at  the  eBOSS
redshifts with respect to the other measurements.

We can  then estimate,  for each reconstruction  $n$ and  method $X\in
\{\dang,H\}$, the weighted average 
\begin{align}
 \mean{\Hord}_{X,n} & =\frac{\one^T \cdot \tens C_n^{-1} \cdot \vect{\Hord}|_{X,n}}{\one^T \cdot \tens C_n^{-1} \cdot \one} ,
\end{align}
where $\vect{\Hord}|_{X,n}$ is
a vector constituted  of estimations of \Hord\  at different redshifts
for iteration $n$, and $\tens{C}_n$ is the associated covariance matrix (due to the correlation in the BAO data).  
We report our results in Table~\ref{tab:Hord}.
The  Planck  2015  value  of  \Hord\ for  the  $\Lambda$CDM  model  is
\SI{9944\pm 127.4}{km.s^{-1}.Mpc^{-1}}. 
We should note an important interpretation of this result.
While all our reconstructions of the expansion history from supernovae
data have  better $\chi^2$ with  respect to the best  fit $\Lambda$CDM
model,  our  large  uncertainties  on  $\Hord$  indicates  that  tight
constraints on this  quantity from model dependent  approaches (such as
assuming  $\Lambda$CDM  model)  have  limitations  in  expressing  the
reality of the universe and estimating its key parameters.

\begin{table}
  \caption{\label{tab:Hord}Weighted average of \Hord\  from the $H$ and \dang\ methods.}
  \centering
  \begin{tabular}{llll}
    \toprule	
    Method & \Hord & Error SN & Error BAO \\
    \midrule
    $\mean{{\Hord}}_{\dang}$ &10120.42  & $^{+33.79}_{-59.12}$  &  $\pm 103.92$\\
    (\si{km.s^{-1}}) &          &                           & \\
    $\mean{{\Hord}}_H$      &  9162.80 & $^{+875.06}_{-921.02}$ & $\pm 166.39 $ \\
    (\si{km.s^{-1}}) &          &                           & \\
    \bottomrule
  \end{tabular}
\end{table}

\subsection{Test of the FLRW metric and the curvature}

\begin{figure}
\centering	
  \includegraphics[width=\columnwidth]{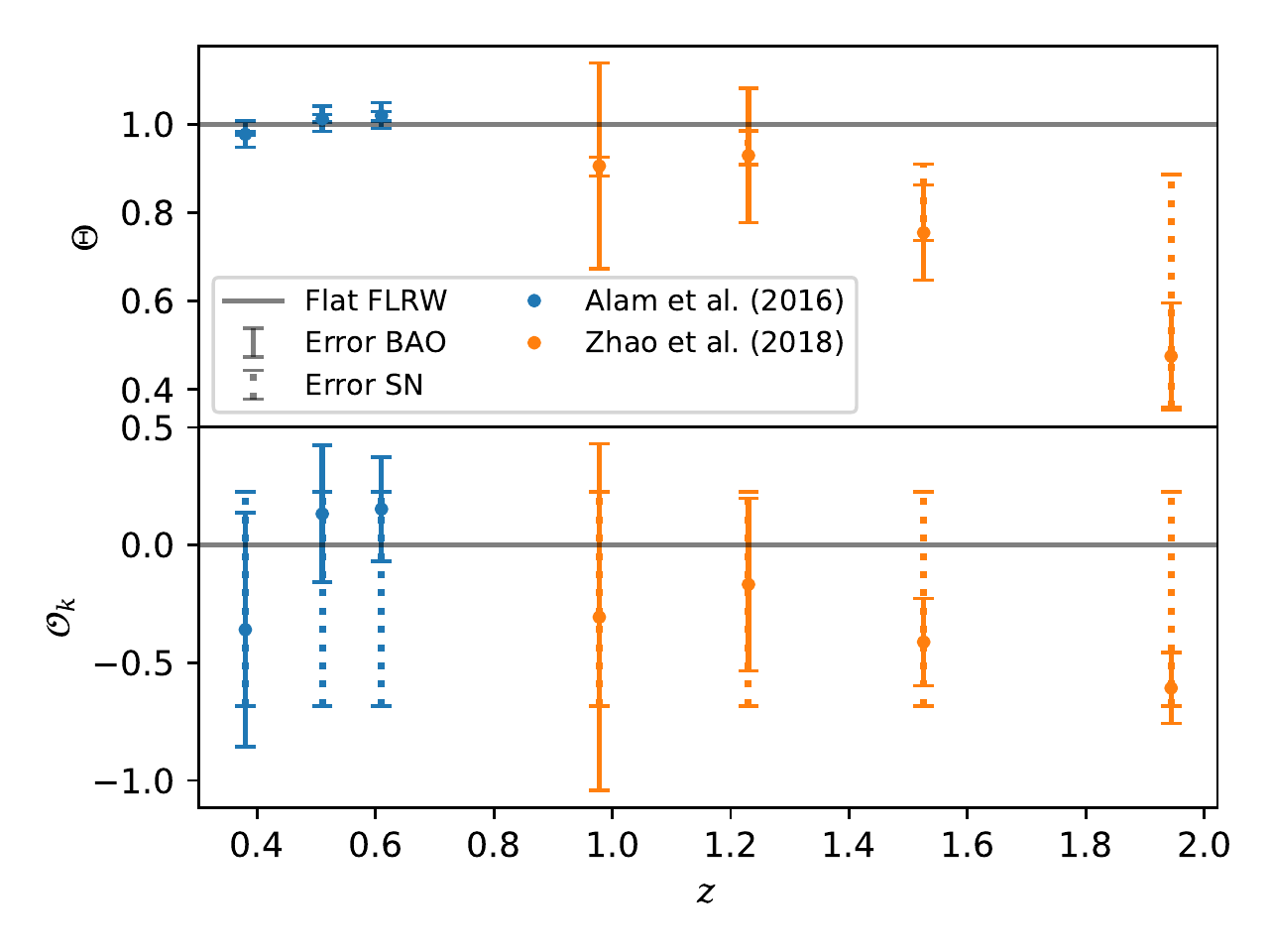}
  \caption{\label{fig:ktest}FLRW  and curvature  test: $\Theta(z)$  (top)
    and $\Ok(z)$ (bottom).
  The dotted error-bars show the range of possible central values from
  different  reconstructions (SN  error), while  the solid  error bars
  show the uncertainty on the central value (BAO error). 
  For a  flat-FLRW Universe, $\Theta(z)  \equiv 1$ and  $\Ok(z) \equiv
  \Omk$.
  } 
\end{figure}

\citet{2017JCAP...01..015L}              reformulated              the
  \Ok\   diagnostic \citep{2008PhRvL.101a1301C}  by  introducing   the
$\Theta$  diagnostic so  that  it  now only  depends  on  the BAO  and
supernovae observables: 
\begin{subequations}
  \begin{align}
  		\label{eq:ktest}
  	\Ok(z) & = \frac{\Theta^2(z)-1}{\DD^2(z)}\\
  	\Theta   (z)    &   =    h(z)\DD'(z)   =   \frac    {1+z}c   H(z)\rd
  	\frac{\dang(z)}{\rd }\frac{\DD'(z)}{\DD(z)}.
  \end{align}
\end{subequations}
For a  FLRW Universe, $\Ok(z) \equiv
  \Omk$, and  in case  of flatness, $\Ok(z)  \equiv 0$  and $\Theta(z)
  \equiv 1$. 
  We can  then calculate  for each  reconstruction $n$  the associated
  $\mathcal{O}_{k,n}(z)$ and $\Theta_n(z)$. 
 We calculated the median of \Ok\ and
  $\Theta$ over all  reconstructions, and defined the SN  error as the
  minimal and maximal  values, and the BAO error as  the maximal error
  over all reconstructions.  
Fig.~\ref{fig:ktest} shows $\Theta(z)$ (top) and $\Ok(z)$ (bottom). 
Both are consistent with a flat FLRW metric up to $z\simeq 1.2$. 

However, at high redshift, some deviation from flatness can be seen. 
Again, this can be explained by the previous remarks. 
In  addition to  the scarcity  of the  SN data  at $z\geq  1.5$, which
results in into poor constraints on  $h(z)$, the BAO seem to show some
internal tensions.
While $\dang(z)/\rd$ are consistent with the Planck best-fit, $H(z)\rd$
are lower than expected.
However, the $\Theta$ and \Ok\  statistics assume a FLRW metric, where
\dang\ and $H$ are related to each other. 
Thus, discrepancy between \dang\ and  $H$ combined with the higher $h$
values at high-redshift  ($z\geq 1$) yields lower  values for $\Theta$
and \Ok.  
{We should also note that in the case of supernova data, the Malmquist bias (if not treated carefully) can pull down the $\DD(z)$ relation at high redshifts. This might explain the large swing upward of $h(z)$ and $\Om(z)$ (with respect to the best-fit \lcdm\ case) that we can see in Fig.~\ref{fig:bao}, and consequently the apparent deviation from flatness observed in $\Theta$ and \Ok.
While it is certainly important to study further this effect in the case of the Pantheon data, it is beyond the scope of this paper. 
}

\section{Growth of Structure Versus Expansion: Testing GR}
\label{sec:growth}
At the  perturbation level, the  cosmological growth of  structure can
also serve as a test of gravity 
\citep{2008PhRvD..77b3504N, 2009JCAP...10..004S, 2010PhRvD..82h2001A,
  2012IJMPD..2150064B, 2013PhRvD..87b3520S, 2014PhRvD..90b3006P,
  2015JCAP...01..004G, 2015PhRvD..91f3009R, 2018MNRAS.475.2122M,
  2017PhRvD..96b3542N, 2018PhRvD..97h3510M, 2017MPLA...3250054S,
  2018arXiv180301337K}.  
In the linear regime, the growth of structure in GR follows 
\begin{align}
  \ddot\delta + 2 H \dot\delta -4\pi G \bar\rho \delta &=  0,
\end{align}
where $\delta = \rho/\bar\rho-1$ is  the density contrast with respect
to the mean density of the Universe $\bar\rho$. 
The growth rate 
\begin{align}
  f(a) & = \deriv {\ln \delta}{\ln a}
\end{align}
can   be    approximated   for    a   wide   range    of   cosmologies
by \citep{1991MNRAS.251..128L, 1998ApJ...508..483W, 2005PhRvD..72d3529L} 
\begin{align}
  f(z) & = \Omm^\gamma(z),
  \label{eq:gamma}
  \intertext{where}
  \Omm(z) & = \frac{\Omm(1+z)^3}{h^2(z)}.
\end{align}
In general relativity (GR), $\gamma\simeq 0.55$.
\fseight\ is thus a powerful probe of gravity.
Observationally,  redshift-space  distortion  enables to  measure  the
combination \citep[e.g.][]{2009JCAP...10..004S} 
\begin{align}
  \label{eq:fseight}
  f\sigma_8(z) & \simeq \sigma_8\Omm^\gamma(z) 
  \exp\left(-\int_0^z\Omm^\gamma(x)\frac{\diff x}{1+x}\right),
\end{align}
where $\sigma_8  = \sigma_8(z=0)$ is  the rms fluctuation  in \hMpc{8}
spheres.   Following   this   formalism,  having   model   independent
reconstructions of the expansion  history and $f\sigma_8(z)$ data, one
can    obtain   constraints    on   \Omm,$\gamma$,    and   $\sigma_8$
\citep{2018MNRAS.476.3263L}. 
Note that we consider the estimated \fseight\ data from BAO surveys as an independent and uncorrelated measurements with respect to the supernova data that we used to reconstruct the expansion history.

Note, however, that one should keep in mind that Eq.~\eqref{eq:gamma} is an approximate fit only. In particular, $\gamma$ may not be exactly constant for quintessence---dark energy modelled by a scalar field with some potential minimally coupled to gravity  \citep[][]{2016JCAP...12..037P}. Still both for $\Lambda$CDM and for quintessence-CDM this fit is good since $\frac{\diff\gamma}{\diff z}$ is small as far as $\Omm$ is not too small, see also \citep[][]{2008PhLB..660..439P}. For modified gravity theories like $f(R)$ gravity, the situation can be different \citep{2009JCAP...02..034G,2010PThPh.123..887M}.

\subsection{Cosmological constraints on \texorpdfstring{$\Omm,\gamma,\sigma_8$}{Om,gamma,sigma8}}

Following  \citet{2018MNRAS.476.3263L},   we  combined   the  Pantheon
compilation with the latest measurements of $\fseight$: 
2dFGRS \citep{2009JCAP...10..004S},
WiggleZ \citep{2011MNRAS.415.2876B},
6dFGRS \citep{2012MNRAS.423.3430B},
 VIPERS \citep{2013MNRAS.435..743D},
the SDSS Main galaxy sample \citep{2015MNRAS.449..848H}, 
2MTF \citep{2017MNRAS.471.3135H}, 
BOSS DR12 \citep{2017MNRAS.465.1757G},
FastSound \citep{2016PASJ...68...38O},  and
eBOSS DR14Q \citep{2018arXiv180103043Z}.
In this section, we assume a flat Universe, therefore
\begin{align}
h(z) = \frac 1 {\DD'(z)}.
\end{align}


{It is worth noting that these measurements, coming from different surveys, were obtained assuming different fiducial cosmologies. 
Therefore, we correct for the fiducial cosmology \cite{2016MNRAS.456.3743A,2017PhRvD..96b3542N}. 
The growth \chisq\ for the $n$th reconstruction $h_n(z)$ and parameters $\vect p = (\Omm,\gamma,\sigma_8)$ is thus given by
}
\begin{align}
  \chisq_{n,\fseight}(\vect p)    &=     \vect{\delta{\fseight}_n}    \cdot
   \tens{C}_{\fseight}^{-1}\cdot \vect{\delta{\fseight}_n}, 
\intertext{%
where $\tens{C}_{\fseight}$ is the growth covariance matrix, and the  $i$th  component of  the residual  vector     $\vect{\delta\fseight}_n$ is 
}
{{\delta\fseight}_n}|_i                 &                 =
            {\frac{h_n(z_i)\DD_n(z_i)}{(1+z_i)H_\text{fid}(z_i)d_\text{A,fid}(z_i)}} \widehat{\fseight}_n(z_i|\vect{p},h_n) \nonumber \\         
           & - {\fseight}|_i. 
\intertext{The total \chisq\ for reconstruction $n$ and parameter \vect{p} is then}
\chi^2_{n,\text{tot}}(\vect{p}) & = \chsq_{n,\fseight}(\vect p)+\chi^2_{n,\text{SN}},
\end{align}
{%
where $\widehat{\fseight}(z_i|\vect p,h_n)$ is the model corresponding to the expansion history $h_n$ and parameters $\vect p$ and fid stands for the fiducial cosmology used by the survey to estimate the data point.  
}



\begin{figure}
     \includegraphics[width=\columnwidth]{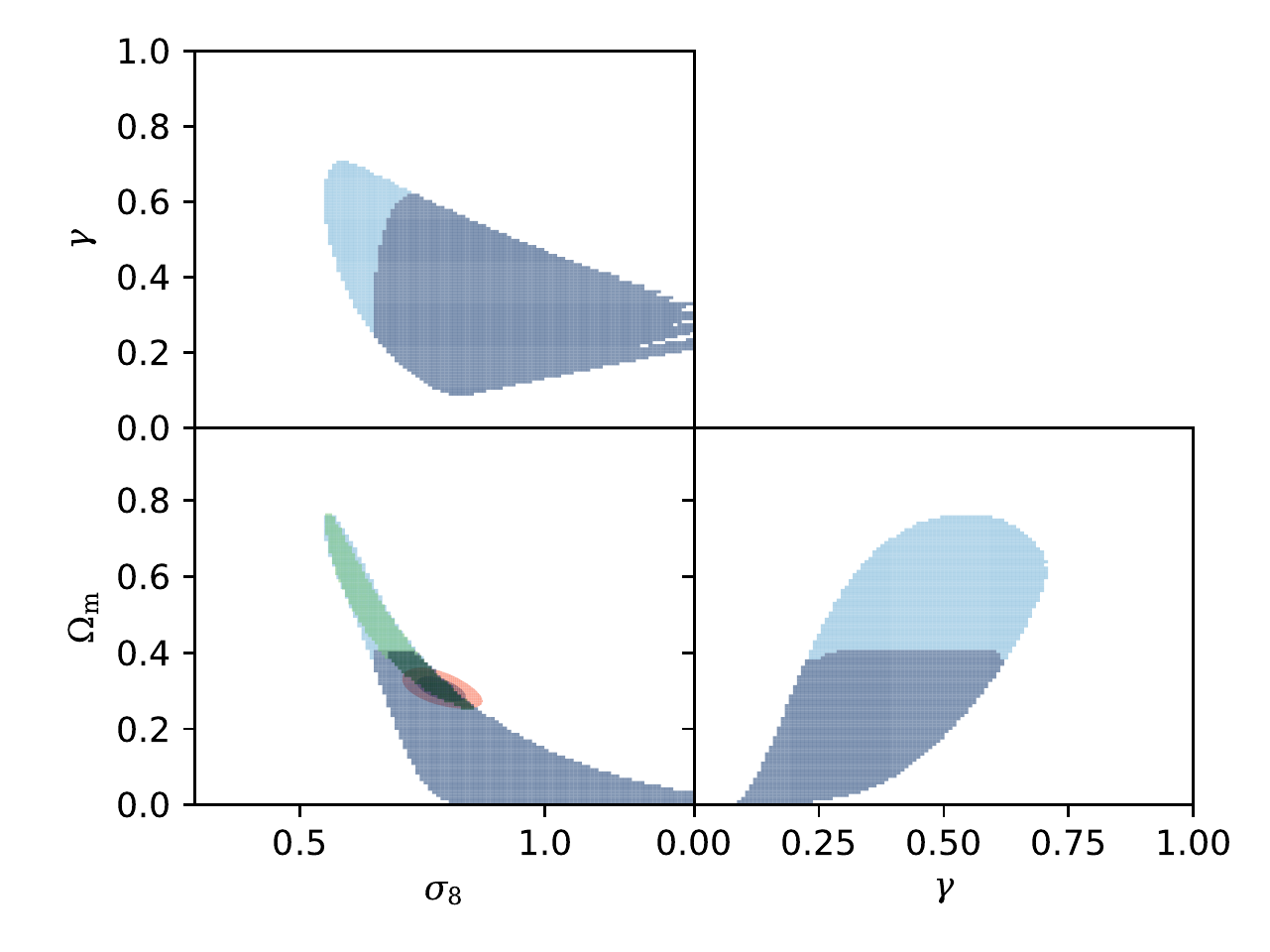}
  \caption{\label{fig:growthexp}    Model    independent    cosmological
    constraints  on $(\Omm,\gamma,\sigma_8)$  from  growth and  expansion
    data.  The  red contours are  the $1\sigma$ and $2\sigma$  confidence levels for
    the \lcdm\ case. The blue contours are associated to the combination
    of the parameters and reconstructions  of the expansion history that
    yield   a   better   \chisq\    with   respect   to   the   best-fit
    \lcdm\  model. The  dark-blue  region satisfy  positive dark  energy
    density condition as expressed in equation~\eqref{eq:posde}. 
   The green contours show the model-independent case where we fixed 
   $\gamma=0.55$, i.e., impose GR. 
   Again, the dark contours satisfy equation~\eqref{eq:posde}.
  }
\end{figure}

The    red    contours    in   the    $(\sigma_8,\Omm)$    plane    in
Fig.~\ref{fig:growthexp} show the $1\sigma$  and $2\sigma$ regions of the
parameter    space   in    the    flat   \lcdm\    case,   that    is,
flat-\lcdm\ expansion history and $\gamma=0.55$. 
The blue contours show the allowed parameter space in the model-independent case.  
Namely, for any point in the blue  contours, one can find at least one
reconstruction   $h(z)$   which,   combined   to   the   corresponding
$(\Omm,\gamma,\sigma_8)$, gives  a better  fit to  the data  than the
best-fit \lcdm.  
In the  $(\sigma_8,\Omm)$ plane,  the model-independent case  is fully
consistent with the \lcdm\ case.  
Moreover, the  flexibility of the model-independent  approach allows a
larger area of the parameter space to be consistent to the data. 
For  instance, for  larger values  of $\sigma_8$  and lower  values of
$\Omm$, one  can find  reconstructed expansion  histories that  give a
better total  fit to the data  (SNIa+growth) with respect to  the best
fit $\Lambda$CDM model.
For  the model-independent  case,  $\gamma$ is  fully consistent  with
0.55, as expected from GR. 
Moreover,  lower  value of  $\gamma$,  combined  with lower  value  of
\Omm\ and larger $\sigma_8$, can also provide good fit to the data. 

We then fix $\gamma=0.55$, as we did  for the \lcdm\ case, and show in
Fig.~\ref{fig:growthexp} the corresponding confidence contours in green. 
This effectively  allows for  a non-\lcdm\ background  expansion, with
gravity as GR. This time, since we  do not allow $\gamma$ to vary, the
region with low \Omm\ and high $\sigma_8$ is now forbidden.  

Finally,   following   \citet{2018MNRAS.476.3263L},    we   focus   on
combinations of $h(z)$ and \Omm\ that respect the positive dark energy
condition 
\begin{align}
\label{eq:posde}
\Omde(z) & = h^2(z) - \Omm(1+z)^3 \geq 0 \qquad \forall z.
\end{align}

We   show  this   region  in   dark-blue (free $\gamma$) and dark-green (fixed $\gamma$) in   Fig.  \ref{fig:growthexp}. 
Imposing equation~\eqref{eq:posde} effectively forbids large values of
$\Omm$, and  dramatically reduces the  allowed parameter space  of the
model-independent case.  
The allowed  region of  the parameter space  is then  fully consistent
with the model-dependent case, as in \citet{2018MNRAS.476.3263L}. 
This is a  strong support from the data for  combination of \lcdm\ and
GR.  Comparing our results here using most recent supernovae (Pantheon
compilation) and BAO data (from eBOSS  DR14) with what was reported in
\citet{2018MNRAS.476.3263L} we  can notice substantial  improvement on
the constraints on these three key cosmological parameters.
Based on  our analysis we  can now put strong  model-independent upper
bound limits  on $\Omm <0.42$  and $\gamma <  0.58$ and a  lower bound
limit on $\sigma_8 > 0.70$.
These  are  in   fact  model  independent  constraints   on  these  key
cosmological parameters.

\section{Summary and Conclusions}

\label{sec:ccl}

We  used  the  Pantheon  supernovae  compilation  to  reconstruct  the
expansion  history  in  a  model-independent way,  using  an  improved
version  of  the  iterative  smoothing method  \citep{2006MNRAS.366.1081S,
  2007MNRAS.380.1573S,2017JCAP...01..015L, 2010PhRvD..81h3537S},  which  we
modified  to take  into account  the  non-diagonal terms  of the  full
covariance matrix.  
We then combined the reconstructed expansion histories to measurements
of  $H(z)\rd$ and  $\dang(z)/\rd$ from  BOSS DR12  and eBOSS  DR14Q to
model-independently measure $H_0\rd$ and test the FLRW metric. 
Our  measurements of  $H_0\rd$  are consistent  with  the Planck  2015
values, while the metric test is consistent with a Flat-FLRW metric.  
However,  for the  eBOSS DR14Q  data points,  while $\dang(z)/\rd$  is
consistent with  the prediction from the  Planck best-fit $\Lambda$CDM
cosmology, the $H(z)\rd$ measurements  are slightly but systematically
lower.  
This   yields   some   hints    for   a   departure   from   flat-FLRW
(Fig.~\ref{fig:ktest})    and    supports    previous    findings    of
\citet[][]{2014ApJ...793L..40S} \& \citet[][]{ 2017NatAs...1..627Z}.

We then  fit the  growth data from  redshift space  distortion, mainly
from SDSS  survey using  the model-independent reconstructions  of the
expansion  history, and  put  model-independent  constraints on  $\Omm
<0.42$, $\gamma < 0.58$ and $\sigma_8 > 0.70$.  
Our measurements  are fully consistent  with the \lcdm\ model  with GR
($\gamma\approx 0.55)$, and  do not reveal  any tension between the  
two data sets.  

Future  surveys,  such as  the  Dark  Energy Spectroscopic  Instrument
\citep{2016arXiv161100036D,2016arXiv161100037D},  the  Large  Synoptic
Telescope  \citep[][]{2008arXiv0805.2366I}, and  WFIRST, will  improve
the quality and  quantity of data, enabling us to  detect any possible
deviation from \lcdm.

\section*{Acknowledgements}

We thank Gongbo Zhao and Yuting Wang for useful discussions, and 
Dan Scolnic for providing the Pantheon data.
This work benefited from the  Supercomputing Center/Korea Institute
of Science  and Technology  Information with  supercomputing resources
including technical support (KSC-2016-C2-0035 and KSC-2017-C2-0021) 
and the  high  performance
computing clusters  Polaris and  Seondeok at  the Korea  Astronomy and
Space Science Institute. A.S. would like to acknowledge the support of
the National Research Foundation of Korea (NRF-2016R1C1B2016478).  
A.~A.~S. was partly supported by the program P-28 ``Cosmos''
of the Russian Academy of Sciences (the project number 0033-2018-0013 of
the Federal Agency of Scientific Organizations of Russia).




\bibliographystyle{apsrev}
\bibliography{biblio} 








\label{lastpage}
\end{document}